\documentclass[twocolumn,showpacs]{revtex4}
\usepackage{graphics}
\usepackage{amsmath}
\usepackage{epsf}
\usepackage{hyperref}
\begin{document}

\title{Thermalization in  mixtures of ultracold gases}
\author{M. Anderlini}
\affiliation{National Institute of Standards and Technology,
Gaithersburg, MD 20899, USA}
\author{D. Gu\'ery-Odelin}
\affiliation{Laboratoire Kastler Brossel, 24, Rue Lhomond, F-75231
Paris Cedex 05, France}

\date{\today}
\begin{abstract}
Starting from a set of coupled Boltzmann equations, we investigate
the thermalization of a two-species cold atomic gas confined
either in a box or in an isotropic harmonic trap. We show that the
thermalization times, by contrast to the collision rate, depend on
the interferences between scattering partial waves. The dynamics
of thermalization in a harmonic trap is also strongly dependent
upon the ratio between the collision rate and the trap
frequencies.
\end{abstract}

\maketitle

\section{Introduction}
Ultracold mixtures of alkali atoms have attracted great interest
among atomic physicists. The ability to simultaneously cool and
trap multiple species offers the possibility to study a large
variety of physical phenomena not previously accessible with a
single species such as novel degenerate quantum systems made of
Fermi-Bose mixtures \cite{fbmixture}.

Thermalization plays a crucial role in the cooling of a sample by
evaporation, and/or in the implementation of sympathetic cooling.
The detailed knowledge of the underlying physics is all the more
important for experiments where the time available to perform
evaporation is limited \cite{lahaye}, or for the optimization of
the evaporation ramp in the multi-partial-wave regime
\cite{doyle}. In addition the return to equilibrium of a gas
initially prepared in an out-of-equilibrium state is a widely used
way to infer the value of the scattering length
\cite{monroe,arndt,delannoy,mosk,rbk,schmidt,anderlini}.

The purpose of this paper is to provide a theoretical framework to
describe the thermalization between two cold gases confined in a
box or a harmonic trap. In contrast to previous theoretical
studies on this subject \cite{delannoy,mosk,kavoulakis}, we
explicitly take into account all partial waves involved in the
collision process, and we work out the role played by the
confinement in all collisional regimes. Since the basic mechanism
described by the Boltzmann equation is binary collisions,
two-component systems exhibit all the complexities of
$n$-component systems. The generalization of the results we obtain
to systems containing more than two species is therefore
straightforward.

The paper is organized as follows. Section \ref{theo} describes
the theoretical framework used to investigate the dynamics of
thermalization. Section \ref{box} is devoted to the simple case of
a gas confined in a box. It gives the explicit expression for the
rethermalization time including the interference terms between all
partial waves involved in the collision process. Section
\ref{harmo} addresses the same problem for a gas trapped in an
isotropic harmonic trap.

\section{Theoretical framework}
\label{theo}

 The gas $i$ ($i=1,2$) consists of $N_i$ particles of mass
$m_i$ initially thermalized at $T^0_i$. The dynamics of the gas is
described by a set of two coupled Boltzmann equations for each
phase-space distribution function $f_i({\bf r}, {\bf v}_i, t)$:
\begin{eqnarray}
\frac{\partial f_i}{\partial
t}+\{f_i,H_i\}&=&I_{ii}[f_i]+I_{ij}[f_i,f_j], \;\; i\neq j.
\label{be}
\end{eqnarray}

The distribution function $f_i$ for the species $i$ is governed by
the hamiltonian $H_i$ through the Poisson bracket $\{f_i,H_i\}$,
and by binary collisions. The appearance of two collision terms on
the right-hand side of Eq. (\ref{be}) is a result of the
possibility of $f_i$ to change through both self collisions
($i$-$i$ collisions) or cross collisions (1-2). The expression for
the collision integrals is
\begin{eqnarray}
&&I_{ij}[f_i,f_j]=\int d^2 \Omega' d^3 v_B |{\bf v}_B-{\bf
v}_A|\frac{d \sigma_{ij}}{d^2 \Omega'} \times \nonumber \\
&&[f_i({\bf r},{\bf v}'_A,t)f_j({\bf r},{\bf v}'_B,t)-f_i({\bf
r},{\bf v}_A,t)f_j({\bf r},{\bf v}_B,t)],\label{intij}
\end{eqnarray}
where $i,j=1,2$. They account for elastic collisions between
particles labelled $A$ and $B$. We denote the velocities of the
two atoms before they interact with each other by  ${\bf v}_A$ and
${\bf v}_B$, and after the collision by ${\bf v'}_A$ and ${\bf
v'}_B$. The expression for the collision terms can be easily
extended to include the effects of both Bose and Fermi statistics
leading to the Boltzmann-Nordheim equation \cite{Nordheim,Baym}.
In fact, most of the results discussed in this paper also hold in
the presence of quantum degeneracy, provided the system is not
Bose-Einstein condensed. The quantitative estimates of collision
effects presented in this work are, however, based on classical
statistics. As shown in Refs \cite{kavoulakis} for bosons and
\cite{vichi} for fermions, our approach, based on the result of
the classical physics, is valid as soon as the temperature of the
sample is larger than twice the temperature for degeneracy, i.e.
the critical temperature for bosons and the Fermi temperature for
fermions.

The collision problem is simplified by an appropriate change of
variables. We introduce the center of mass velocity ${\bf
v}_0=(m_A {\bf v}_A+m_B{\bf v}_B)/(m_A+m_B)$, the total mass
$M=m_A+m_B$, the relative velocity ${\bf v}_r={\bf v}_A-{\bf
v}_B$, and the reduced mass $\mu=m_Am_B/M$. The relative velocity
changes from the value ${\bf v}_r$ before the collision to the
value ${\bf v}'_r$ after the collision, keeping the same value for
the modulus $v_r=v'_r$ because of energy conservation. The
collision process is described by specifying the scattering
direction with the polar angle $\theta'$ and the azimuthal angle
$\varphi'$ of the final relative velocity ${\bf v}'_r$ with
respect to the initial relative velocity ${\bf v}_r$ before the
collision. The corresponding solid angle $d^2\Omega'=d\varphi'
\sin \theta'd\theta'$ enters the expression for the integrand of
the collision integrals.

The angular dependence of the elastic scattering directly derives
from the partial-waves decomposition of the scattering amplitude
in the quantum mechanical description of interactions
\cite{joachain}. For non-identical atoms, the asymptotic form of
the scattering amplitude $f(k,\theta')$ reads
\begin{equation}
f(k,\theta')=\frac{1}{k}\sum_{l=0}^\infty(2l+1)e^{i\delta_l}\sin(\delta_l)P_l[\cos({\theta')}],
\label{ad}
\end{equation} where $k=\mu v_r/\hbar$,
$\theta'\in[0,\pi)$  and $P_l[\cos({\theta')}]$ are the Legendre
polynomials. All the complexity of the interatomic potential is
contained in the $k$-dependence of the phase shifts $\delta_l$.
The differential cross section for non-identical atoms is given by
$d\sigma_{12}/d\Omega'=|f(k,\theta')|^2$.

For identical atoms, we have to take into account the
(anti)symmetrization principle. Accordingly the differential cross
section takes the form:
\begin{equation}
\frac{d\sigma_{ii}}{d\Omega'}=|f(k,\theta')+\epsilon
f(k,\pi-\theta')|^2,\label{csi}
\end{equation}
where $\epsilon=1$ for bosons and $\epsilon=-1$ for fermions. The
integration must be carried out only in the half sphere
$\theta'\in[0,\pi/2)$. As a consequence of the parity of the
spherical harmonic, the only partial waves contributing to the
scattering cross-section for polarized bosons (respectively,
fermions) correspond to even (respectively, odd) values of $l$.
The interference between partial waves are contained in the
differential cross section \cite{jookthomas}.

A relevant parameter to describe the thermalization of the gases
is the collision rate. The average total number of collisions per
unit of time $\Gamma_{ij}$ for atom species $i$ and $j$ is
obtained by integrating over space and velocity the output channel
term of the collision integral in Eq. (\ref{intij}):
\begin{equation}
\Gamma_{ij}=\int d^3rd^3v_Ad^3v_B \sigma_{ij}|{\bf v}_A-{\bf
v}_B|f_i({\bf r}, {\bf v}_A,t)f_j({\bf r}, {\bf v}_B,t)
\label{Numberij}
\end{equation}
 From this quantity we deduce the expression for the mean
collision rate per atom $\gamma_{ij}= \Gamma_{ij}/N_i$. The
integrated cross section $\sigma_{ij}$ is just the sum of the
contributions from partial-waves:
\begin{equation}
\sigma_{ij}(v_r)=(1+\delta_{ij})\sum_l
\frac{4\pi}{k^2}(2l+1)\sin^2\delta_l, \label{sij}
\end{equation}
where the factor $(1+\delta_{i,j})$ accounts for the constructive
interference of undistinguishable scattering channels for
identical atoms, $\delta_{i,j}$ being the Kronecker delta symbol.
The sum in Eq. (\ref{sij}) is taken over even (resp. odd) values
for identical ($i=j$) bosons (resp. fermions), and over all
integer values for non-identical atoms ($i\neq j$). The expression
for the collision rate is just a simple sum over the partial
waves, it does not exhibit interference. This is to be contrasted
with the expression of the thermalization rate as will be
illustrated in the next section.

The determination of the thermalization rate is based on the
method of averages \cite{dgoPRA99} combined with an appropriate
ansatz. Starting from Eq. (\ref{be}), one readily derives the
equation for the average value of a general dynamical quantity
$\mathcal{O}({\bf r},{\bf v}_i)$:
\begin{equation}
\frac{d \langle \mathcal{O}\rangle_i }{dt} + \langle\mathcal{O}
\{f_i,H_i\} \rangle_i = \langle \mathcal{O}
I_{ii}\rangle_i+\langle\mathcal{O} I_{ij}\rangle_i,
\label{momenteq}
\end{equation}
where the average is taken in both position and velocity space: $
\langle \mathcal{O} \rangle_i = \int d^3 r d^3 v_i f_i({\bf r},
{\bf v}_i, t) \mathcal{O}({\bf r}, {\bf v}_i)/N_i$. As a
consequence of the collisional invariants --- number of atoms,
momentum, and energy ---, $\langle \mathcal{O}_i I_{ii}\rangle_i =
0$ if $\mathcal{O}_i$ is of the form $\mathcal{O}_i = a({\bf
r})+{\bf b}({\bf r})\cdot {\bf v}_i+c({\bf r}){\bf v}^2_i$.

\section{Thermalization in a box}
\label{box}

As a first application of the model, we consider the
thermalization of a two-species gas in a box of volume $V$. We
assume that atoms can undergo only specular reflection on the
walls of the box, which can be realized experimentally by using
the dipolar optical force \cite{raizen}. Initially, the gases are
at different temperatures $T^0_1\neq T^0_2$. After thermalization,
the temperature of the gases will be the same $T_f = (N_1
T^0_1+N_2 T^0_2)/(N_1+N_2)$, within the Boltzmann dynamics. To
evaluate the thermalization time, we write the equation for the
mean total kinetic energy $\langle N_im_i{\bf v^2}_i/2\rangle_i$
of species $i$:
\begin{equation}
\frac{d}{dt}\left\langle \frac{N_i m_i{\bf
v}^2_i}{2}\right\rangle_i=\left\langle\frac{N_i m_i{\bf v}^2_i}{2}
I_{ij}\right\rangle_i\equiv \Sigma_i,\;\;j\neq i.\;\;
\label{eqbox}
\end{equation}
The Poisson bracket of Eq. (\ref{momenteq}) vanishes for a
homogeneous system, and the self collision integral term does not
contribute because of conservation of kinetic energy for elastic
collisions. The calculation of the collision integral term
requires an ansatz for the distribution function. We choose a
Gaussian ansatz for the phase-space distribution function of each
species:
\begin{equation}
 f_i({\bf v_i},t) = \mathcal{N}_i\exp\left(-\frac{m_i v^2_i}{2
 k_BT_i}\right), \label{gaussian}
\end{equation}
where $\mathcal{N}_i=N_im_i^{3/2}/(V(2\pi k_BT_i)^{3/2})$ is the
normalization factor. The time dependence is contained only in the
effective temperatures $T_i$.  We assume that $(T^0_1-T_f )/T_f
\ll 1$ and $(T^0_2-T_f )/T_f \ll 1$. Consequently, we can evaluate
the total number of cross collisions per unit of time by setting
$T_1=T_2=T_f$ in Eq. (\ref{Numberij}) with the Gaussian ansatz for
the distribution functions:
\begin{eqnarray}
\Gamma_{12} =  \frac{2}{\sqrt{\pi}}\frac{N_1 N_2}{V}c \langle
\sigma_{12}\rangle, \label{numberbox}
\end{eqnarray}
where we have introduced the velocity $c=(2k_BT_f/\mu)^{1/2}$ and
the thermally averaged cross section $\langle \sigma_{12}\rangle=
2\int_0^\infty dx\, \sigma_{12}(c x) x^3 e^{-x^2}$.

The expansion around the final temperature $T_f$ of the collision
terms yields a set of linear equations. We can rewrite them in the
form (see Appendix \ref{A}):
\begin{equation}
\frac{d(T_i-T_j)}{dt}=-\left(
\frac{N_i+N_j}{N_iN_j}\right)\frac{T_i-T_j}{\tau}\;,\;i\neq j,
\label{dtbox}
\end{equation}
with
\begin{equation}
\frac{1}{\tau}=\frac{8\mu}{3M}\frac{\langle\!\langle \sigma_{12}
\rangle\!\rangle}{\langle \sigma_{12} \rangle}\Gamma_{12},
\label{taubox}
\end{equation}
where $\langle\!\langle \sigma_{12}\rangle\!\rangle= \int_0^\infty
dx\, \tilde{\sigma}_{12}(c x) x^5 e^{-x^2}$. The quantity
$\tilde{\sigma}_{12}(c x)$ is defined as:
\begin{equation}
\tilde{\sigma}_{12}(c x)=2\pi\int_0^\pi
\sin\theta'(1-\cos\theta')\frac{d\sigma_{12}}{d\Omega'}d\theta'.
\label{st12}
\end{equation}
The new feature of this expression arises from the interference
between partial waves resulting from the angular integration. This
is in contrast to the expression for the total cross section,
which just contains a simple sum over partial waves. We show in
Appendix \ref{A} that interference occur only between partial
waves differing by at most one unit of angular momentum.

We thus find that the relaxation corresponds to an exponential
decay but the thermalization times $\tau$ is {\it not} in general
proportional to the collision rate because it contains partial
wave interferences terms.

At very low temperature where only $s$-waves contribute to the
collision process, no interference can occur and the relaxation
time is proportional to the inverse of the collision rate. If the
cross section depends on the relative wave vector $k$ of the
collision, the factor of proportionality depends on temperature.

For a cross section of the form $\sigma_{12}(k)=4\pi
a_{12}^2/(1+a_{12}^2k^2)$,  one finds
\begin{eqnarray}
 \langle \sigma_{12}\rangle&=&
4\pi a_{12}^2 \xi [1-\xi e^{\xi}\Gamma(0,\xi)],\nonumber \\
\langle\!\langle\sigma_{12}\rangle \!\rangle & = &  2\pi a_{12}^2
\xi [1-\xi+\xi^2 e^{\xi}\Gamma(0,\xi)],
\end{eqnarray}
where $\xi=\hbar^2/(2\mu k_B T_f a_{12}^2)$ and
$\Gamma(a,z)=\int_z^\infty t^{a-1}e^{-t}dt$. In the very low
temperature limit $\xi\gg 1$, we have $\langle \sigma_{12}\rangle
= \langle\!\langle \sigma_{12}\rangle \!\rangle= 4\pi a_{12}^2$;
in the unitary limit $\xi\ll 1$, we obtain $\langle
\sigma_{12}\rangle = 2\langle\!\langle \sigma_{12}\rangle
\!\rangle=2\pi \hbar^2/(\mu k_B T_f)$. As a consequence, the
number $(\tau\Gamma_{12})$ of inter-species collisions per atom
required to equilibrate the temperature of a two-component system
made of atoms of the same mass varies from 1.5 in the very low
temperature limit to 3 in the unitary limit according to Eq.
(\ref{taubox}).

\begin{figure}[h!]
\centering\begin{center} \mbox{\epsfxsize 3.4 in
\epsfbox{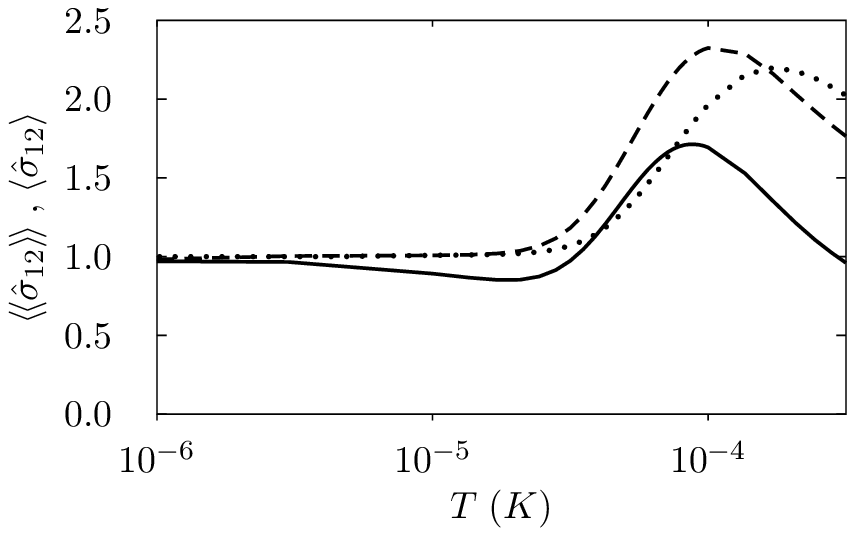}} \caption{Averaged cross sections
$\langle\!\langle \sigma_{12}\rangle\!\rangle$ and $\langle
\sigma_{12}\rangle$ normalized to their value at zero temperature
as a function of the final temperature of the mixture made of
rubidium 87 in the two different internal states
$|5S_{1/2},F=2,m=1\rangle$ and $|5S_{1/2},F=1,m=-1\rangle$. The
resonance shape results from a $d$-wave resonance. The quantity
$\langle \sigma_{12}\rangle$ is represented in dotted lines. The
calculation of $\langle\!\langle \sigma_{12}\rangle\!\rangle$ made
with interference terms (solid line) differs notably from the
calculation where those terms are neglected (dashed
line).}\label{figure1}
\end{center}
\end{figure}

In the following, we explicitly derive the curves for the
thermalization time $\tau$ and the inter-species collision rate
$\gamma$ for the same number of $^{87}$Rb atoms $N_1=N_2=N$ in two
different internal states $|5S_{1/2},F=2,m=1\rangle$ and
$|5S_{1/2},F=1,m=-1\rangle$. This calculation has been performed
by taking into account the first five partial wave contributions
for the cross collisions \cite{servaas}. We plot in Fig.
(\ref{figure1}) the average cross section $\langle\!\langle
\sigma_{12}\rangle\!\rangle$ normalized to its value with the
cross section taken at zero energy. The resonance shape results
from a $d$-wave resonance, which in turn tends to magnify the
interference terms  between $p$-$d$, and $d$-$f$ partial waves,
leading to a significant difference between the thermalization
times calculated with and without the inclusion of interference
terms.

\section{Thermalization in an isotropic harmonic trap}
\label{harmo}

We have shown in the previous section that the relation between
the collision rate and the thermalization rate is strongly
affected by partial wave interferences. In this section we
emphasize the role played by the confinement, and we derive the
dynamics of the thermalization in all collisional regime
\cite{walraven}.

For the sake of simplicity, we consider the thermalization of an
atomic mixture confined in a harmonic and isotropic trap:
\begin{equation}
U_i(x, y, z) = \frac{1}{2} m_i \omega_{i}^2 (x^2+y^2+z^2).
\label{HarmonicPot-Eq}
\end{equation}
By application of the method of averages, we derive the following
set of six coupled equations starting from the evolution of the
square radius:
\begin{eqnarray}
\frac{d\langle {\bf  r}^2\rangle_i}{dt}&=&
2\langle {\bf r}\cdot{\bf v}_i\rangle_i, \nonumber\\
\frac{d\langle {\bf r}\cdot{\bf v}_i\rangle_i}{dt}&=&\langle {\bf
v}^2_i\rangle_i-\omega_{i}^2 \langle {\bf  r}^2\rangle_i +
\langle{\bf r}\cdot{\bf v}_i I_{ij}\rangle_i,\nonumber\\
\frac{d\langle {\bf v}^2_i\rangle_i}{dt}&=&-2\omega_{i}^2 \langle
{\bf r}\cdot{\bf v}_i\rangle_i + \langle {\bf v}^2_i I_{ij}
\rangle_i.\label{LinSystem-Eq}
\end{eqnarray}
The momentum and kinetic energy equations for the individual
species include a contribution from the inter-species collision
term, expressing the fact that it is the total momentum and total
kinetic energy of the system rather than the individual species'
momentum and kinetic energy that are summational invariants. These
dissipative contributions to the species equations do not appear
in the equation for the radius since the collision is local and
the number of atoms for each species is conserved.

In the absence of inter-species interactions the set of equations
(\ref{LinSystem-Eq}) describes the monopole mode
\cite{cercignani88,boltzmann09,dgoPRA99} of both species
independently. The collision terms involving $I_{ii}$ do not
contribute to the above equations because all the dynamic
quantities are collision invariant for intra-species collisions.
This is why there is no damping for the \emph{breathing} mode of a
single-species classical gas confined in a harmonic isotropic
trap. In this particular geometry and in the presence of two
species, the relaxation only occurs through the inter-species
collisions whatever the collisional regime of a given species may
be.

The total number of inter-species collisions per unit of time
$\Gamma_{12}$ can be evaluated after the thermalization, once
equilibrium has been reached:
\begin{equation}
\Gamma_{12}=\frac{N_1N_2c\langle\sigma_{12}\rangle}{\pi^2\sqrt{2}}
\left(\frac{\mu\omega_1\omega_2}{k_BT_f}.
\frac{(m_1+m_2)\omega_1\omega_2}{m_1\omega_1^2+m_2\omega_2^2}\right)^{3/2}.
\label{nh}
\end{equation}
As the initial state of the mixture is assumed to be not too far
from the final state, this expression gives a reliable estimate of
$\Gamma_{12}$ during the thermalization process. The set of Eqs.
(\ref{LinSystem-Eq}) is exact within the Boltzmann formalism.
  To calculate quantitatively the
relaxation driven by inter-species collisions, we make a Gaussian
ansatz for the phase-space distribution function $f_i\equiv
f_i({\bf r},{\bf v}_i,t)$ of each species, with the inclusion of a
factor taking into account the space-velocity correlations:
\begin{equation}
f_i=\mathcal{N}_i\exp\left(-\frac{m_iv_i^2+m_i\omega_i^2r^2}{2k_BT_i}\right)(1+\eta_im_i{\bf
r}.{\bf v}_i), \label{ansatzh}
\end{equation}
where $\mathcal{N}_i$ is the normalization factor. The effective
temperature $T_i$ and the parameter $\eta_i$ are the only
time-dependent variables. Such an ansatz is inspired by the exact
solution for the phase-space distribution function for the
monopole mode, and provides a natural generalization of the local
equilibrium distribution. For one species the Gaussian ansatz was
shown to be accurate for investigating the damping of the coupled
monopole-quadrupole oscillations in an anisotropic harmonic trap
from the collisionless to the hydrodynamic regime \cite{dgoPRA99}.
At the lowest order, the space-velocity correlations only enter
the momentum equations $\langle{\bf r}\cdot{\bf
v}_iI_{ij}\rangle_i$, while they can be neglected for the
calculation of the term involving the kinetic energy $\langle{\bf
v}_i^2I_{ij}\rangle_i$. The details of this calculation can be
found in Appendix \ref{B}. We find:
\begin{eqnarray}
\langle{\bf r}\cdot{\bf v}_1I_{12}\rangle_1 & = &  - \frac{m_1
\omega_1^2\langle {\bf r}\cdot{\bf v}_1\rangle_1-
m_2\omega_2^2\langle {\bf r}\cdot{\bf v}_2 \rangle_2}{
m_1N_1\omega_1 \omega_2\,\tilde{\tau}},\label{RelaxR.V-Eq}
\\ \langle{\bf v}_1^2I_{12}\rangle_1 & = & -\frac{m_1\langle {\bf
v}^2_1\rangle_1- m_2\langle {\bf
v}^2_2\rangle_2}{m_1\,N_1\,\tau},\label{RelaxV2-Eq}
\end{eqnarray}
where the time constants $\tilde{\tau}$ and $\tau$ are given by:
\begin{eqnarray}
\frac{1}{\tilde{\tau}}&=& \frac{4}{3}
\frac{\mu\omega_1\omega_2}{m_1 \omega_1^2 +m_2
\omega_2^2}\frac{\langle\!\langle\sigma_{12}\rangle\!\rangle}{\langle\sigma_{12}\rangle}\Gamma_{12},  \label{tau'12-Eq}\\
\frac{1}{\tau}&=&  \frac{8}{3} \frac{\mu}{M}
\frac{\langle\!\langle\sigma_{12}\rangle\!\rangle}{\langle\sigma_{12}\rangle}\Gamma_{12}.
\label{tau12-Eq}
\end{eqnarray}
Combining Eq. (\ref{LinSystem-Eq}), and Eqs.
(\ref{RelaxR.V-Eq})-(\ref{tau12-Eq}), we obtain a closed set of
six linear coupled equations. As a consequence, the relaxation in
a trap does not correspond in general to a simple exponential
decay.

We demonstrate the physics of rethermalization using a specific
example in which $m_1=m_2=m$, $\omega_1=\omega_2=\omega$,
$N_1=N_2=N$ and keeping a constant cross section. However, we
emphasize that the conclusions we draw are generic. In order to
follow the thermalization it is convenient to introduce the three
quantities:
\begin{eqnarray}
\Delta_1(t)&=&\langle {\bf r}^2\rangle_1(t)-\langle {\bf
r}^2\rangle_2(t),\nonumber\\
\Delta_2(t)&=&\langle {\bf r}\cdot{\bf v}_1\rangle_1(t)-\langle
{\bf
r}\cdot{\bf v}_2\rangle_2(t),\nonumber\\
\Delta_3(t)&=&\langle {\bf v}^2_1\rangle_1(t)-\langle {\bf
v}^2_2\rangle_2(t).\nonumber
\end{eqnarray}

\begin{figure}
\centering\begin{center} \mbox{\epsfxsize 3.2 in
\epsfbox{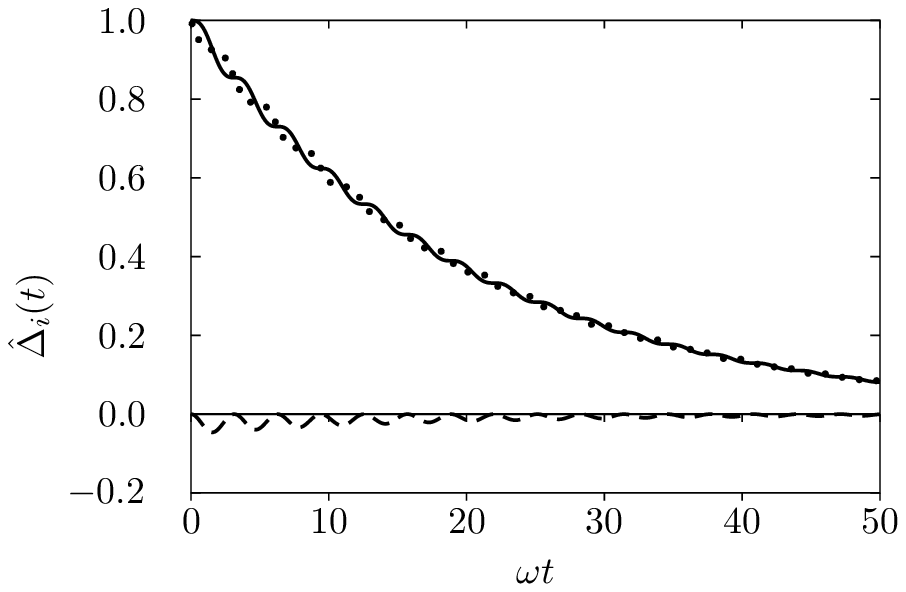}} \caption{Evolution of the normalized moments
$\hat{\Delta}_i$ as a function of $\omega t$ in the collisionless
regime $\omega\tau_0=20$: $\hat{\Delta}_1=\Delta_1(t)/\Delta_1(0)$
(solid line),
$\hat{\Delta}_2=2\Delta_2(t)/(\Delta_1(0)+\Delta_3(0))$ (dashed
line) and $\hat{\Delta}_3=\Delta_3(t)/\Delta_3(0)$. They account
respectively for the difference in the mean value of the square of
the radii, the space-velocity correlations and the square of the
velocity.}\label{figh1}
\end{center}
\end{figure}

\begin{figure}
\centering\begin{center} \mbox{\epsfxsize 3.2 in
\epsfbox{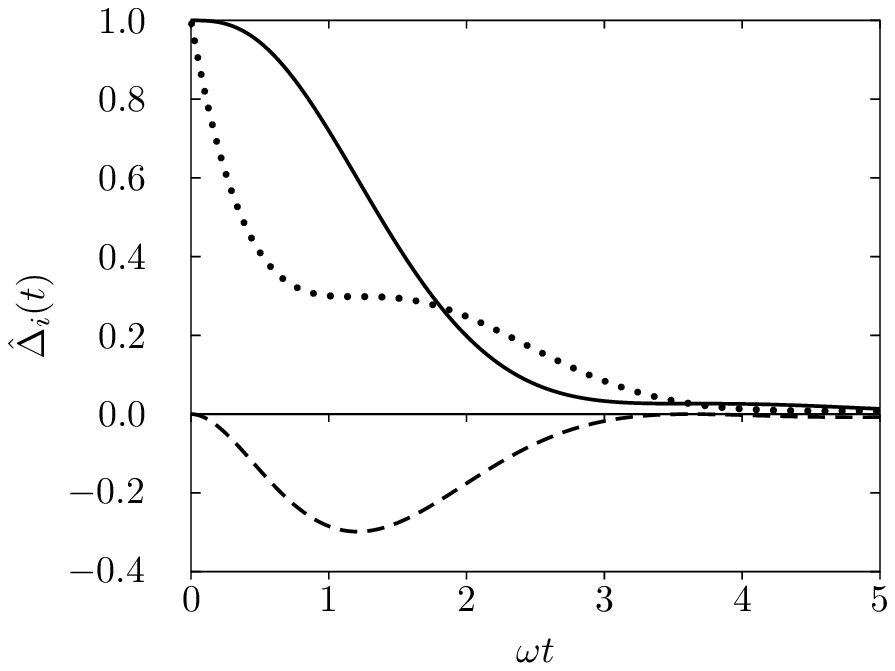}} \caption{Evolution of the normalized moments
$\hat{\Delta}_i$ as a function of $\omega t$ in the intermediate
regime $\omega\tau_0=1$, same notation as in Fig.
(\ref{figh1}).}\label{figh2}
\end{center}
\end{figure}

\begin{figure}
\centering\begin{center} \mbox{\epsfxsize 3.2 in
\epsfbox{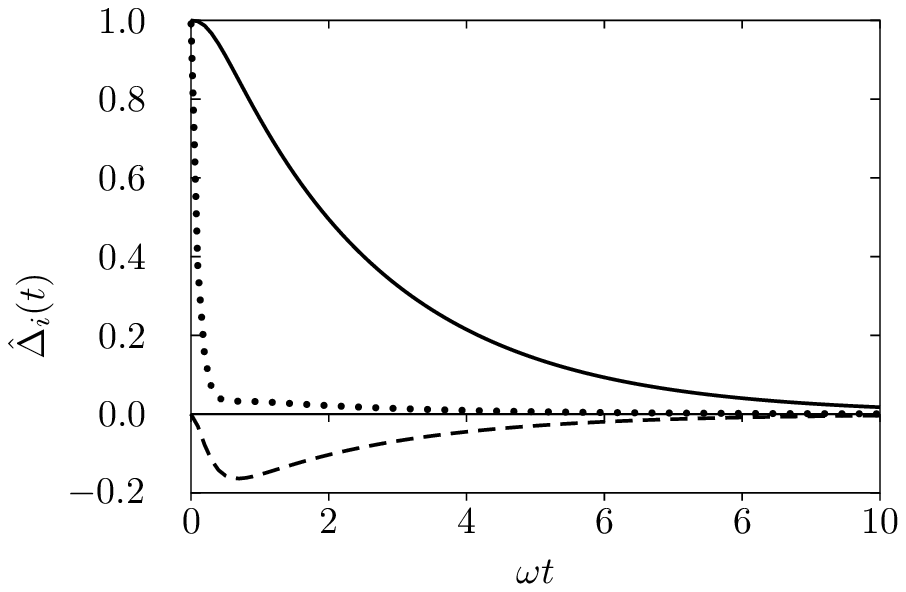}} \caption{Evolution of the normalized moments
$\hat{\Delta}_i$ as a function of $\omega t$ in the hydrodynamic
regime $\omega\tau_0=0.2$, same notation as in Fig.
(\ref{figh1}).}\label{figh3}
\end{center}
\end{figure}

We characterize the collisional regime through the quantity
$\omega\tau_0$ with $\tau_0=N\tau$. The behavior we obtain is
reminiscent of the one of the monopole-quadrupole mode relaxation
in a harmonic and anisotropic trap \cite{dgoPRA99}. However, the
observables  $\langle {\bf r}^2\rangle_i$, $\langle {\bf
r}\cdot{\bf v}_i\rangle_i$ and $\langle {\bf v}^2\rangle_i$ and
their equation of evolution are not the same. The different
regimes depend on the dimensionless parameter $\omega\tau_0$.

For $\omega\tau_0\gg 1$, the gas is in the collisionless regime
(see Fig. \ref{figh1}). We can recover {\it formally} the results
obtained in the previous section, for a confinement of the gas in
a box, by neglecting the space-velocity correlation term in the
equation for $\Delta_3$, leading to an effective equation for
thermalization that is identical to Eq. (\ref{dtbox}). In presence
of the harmonic confinement, $\Delta_3$ obeys a third order
differential equation according to the set of equations
\ref{LinSystem-Eq}. In the collisionless regime, it can be recast,
under our assumptions, in the form:
$$
\frac{d\Delta_3}{dt}\simeq -\frac{\Delta_3}{\tau}.
$$
The time needed for thermalization is longer for a confined
gas(see for comparison Eq. (\ref{dtbox})). This is in agreement
with the results of references \cite{delannoy,mosk}. This factor
of two accounts for the fact that the space of configuration is
larger for a non homogeneous gas, and the thermalization affects
both the space and velocity degrees of freedom.

The decay exhibits oscillations when the collisional regime is
such that $\omega\tau_0
> 0.5$, with an amplitude scaling as $1/\omega\tau_0$.
In the intermediate regime ($\omega\tau_0\sim 1$), the
space-velocity correlations play a crucial role in the dynamics,
and
 the decay of the $\Delta_i$ quantities is not
exponential at all as shown in Fig. \ref{figh2}. In the
hydrodynamic regime ($\omega\tau_0\ll 1$), we find two separate
timescales: a rapid one for the relaxation in velocity space, and
a slow one for the relaxation of the spatial widths (see Fig.
\ref{figh3}). We recover in this limit the classical behavior of
hydrodynamics, with a short time needed to reach a local
equilibrium and a longer time to reach the global equilibrium. The
time needed to reach complete thermalization is proportional to
$1/(\omega^2\tau_0)$. In contrast to the collisionless regime,
this time increases with the inter-species collision rate. We have
also performed numerical simulations based on molecular dynamics
\cite{numsim} to analyse the rethermalization in all collisional
regimes and for arbitrary values of atom numbers, atomic masses,
and harmonic trap frequencies for the two species. We have
obtained a good agreement with the predictions of our model based
on the ansatz (\ref{ansatzh}).

In an anisotropic trap with different trap frequencies
$\omega_{i}^{(x,y,z)}$, one can derive a set of equations similar
to Eq. (\ref{LinSystem-Eq}). In this case the behavior of the
relaxation is also influenced by the damping of the quadrupole
modes \cite{dgoPRA99}.

\section{Conclusion}

In this article, we have investigated the thermalization of a non
degenerate atomic mixture. In our analysis we have taken into
account both the energy and angular dependence of the
inter-species elastic differential cross section. We have derived
the explicit form for the thermalization rate, which depends on
the interference between the partial waves while the collision
rate does not. In addition, we have shown how the dynamics of the
thermalization is modified by the collisional regime in presence
of a confinement.

\section*{Acknowledgements}

We acknowledge S. Kokkelmans for providing us with the phase
shifts in rubidium collisions. We thank D. Roberts for a careful
reading of the manuscript. This work was supported by the
D\'el\'egation G\'en\'erale de l'Armement (DGA).

\appendix

\section{Collision integrals calculation in the homogeneous case}
\label{A}

The global conservation of kinetic energy in a collision leads to
$\Sigma^+=\,\Sigma_1+\Sigma_2=0$. The symmetries of the collision
integral permit one to rewrite the quantity
$\Sigma^-=\,\Sigma_1-\Sigma_2$ in the useful form:
\begin{equation}
\Sigma^-=\mu V \int
d^3v_0d^3v_rd^2\Omega'v_r\frac{d\sigma_{12}}{d\Omega'}\left[{\bf
v}_0\cdot ({\bf v}_r-{\bf v}'_r)\right] \Xi_{1,2}^{1',2'},
\end{equation}
where we have introduced the quantity $\Xi_{1,2}^{1',2'} =f_1({\bf
v}'_1,t)f_1({\bf v}'_2,t)-f_1({\bf v}_1,t)f_1({\bf v}_2,t)$. The
Gaussian ansatz (\ref{gaussian}) for the distribution functions
$f_i$ allows for an expansion of $\Xi_{1,2}^{1',2'}$ around the
final temperature $T_f$. We obtain
\begin{eqnarray}
\Sigma^- =-\frac{\mu^2V}{k_BT_f}\left(
\frac{T_1-T_2}{T_f}\right)\mathcal{N}_1\mathcal{N}_2 \int
d^3v_0d^3v_rd^2\Omega' \nonumber \\
v_r\frac{d\sigma_{12}}{d\Omega'}\left[{\bf v}_0. ({\bf v}_r-{\bf
v}_r')\right]^2\exp\left(-\frac{ M v^2_0+\mu v^2_R}{2 k_B
T_f}\right). \label{XI-Eq}
\end{eqnarray}
From the calculation of $\Sigma^-$, one obtains the set of linear
Eqs. (\ref{dtbox}) with the explicit form Eq. (\ref{taubox}) for
the relaxation times $\tau$. The angular integration of Eq.
(\ref{XI-Eq}) leads to the quantity $\tilde{\sigma}_{12}(c x)$
defined in Eq. (\ref{st12}). From the expansion of the scattering
amplitude in terms of partial waves (\ref{ad}), we can recast it
in the form
\begin{equation}
\tilde{\sigma}_{12}(c x)=\frac{2\pi}{k^2}\sum_{0\leq l \leq l'<
\infty
}\alpha_{l,l'}\sin\delta_l\sin\delta_{l'}\cos(\delta_l-\delta_{l'}),
\label{stpw}
\end{equation}
where we have introduced the dimensionless numerical factors
$\alpha_{l,l'}=(2-\delta_{l,l'})(2l+1)(2l'+1)\int_{-1}^1(1-x)P_l(x)P_{l'}(x)$,
$\delta_{l,l'}$ being the Kronecker delta symbol. The expression
for the coefficients $\alpha_{l,l'}$ explicitly contains terms
describing interference between different partial waves. The
properties of the coefficient $\alpha_{l,l'}$ are the following:
$\alpha_{l,l}>0$, $\alpha_{l,l\pm 1}<0$, $\alpha_{l,l'}=0$
otherwise.

Finally, we perform the Gaussian integration of Eq. (\ref{XI-Eq}),
and we work out the expression for the relaxation time $\tau$ of
Eq. (\ref{taubox}) according to the notation of Eq. (\ref{dtbox}).

\section{Collision integrals calculation for a harmonic confinement}
\label{B}

The calculation of the quantity $\langle {\bf v}_1^2I_{12}\rangle$
can be carried out in a way similar to the one used in Appendix
\ref{A}. The terms in $\eta_i$ of the ansatz (\ref{ansatzh}) do
not contribute at the lowest order.

To calculate the quantity $\langle {\bf r}\cdot{\bf
v}_1I_{12}\rangle$, we introduce the quantities
$\Lambda_i=m_iN_i\langle {\bf r}\cdot{\bf v}_iI_{ij}\rangle_i$ and
$\Lambda^\pm=\Lambda_1\pm\Lambda_2$. The global conservation of
momentum ensures $\Lambda^+=0$. The first non-vanishing
contribution to the linearized expansion of $\Lambda^-$ is
obtained by setting $T_1=T_2=T_f$ in the ansatz (\ref{ansatzh}).
One finds:
\begin{eqnarray}
\Lambda^- =-(\eta_1-\eta_2)\mu^2\mathcal{N}_1\mathcal{N}_2 \int
d^3v_0e^{- M v^2_0/2 k_B T_f}\nonumber \\ \int
d^3rd^3v_rd^2\Omega' v_r\frac{d\sigma_{12}}{d\Omega'}\left[{\bf
r}\cdot({\bf v}_r-{\bf v}'_r )\right]^2 \nonumber
\\\exp\left(-\frac{ (m_1\omega_1^2+m_2\omega_2^2)r^2+\mu v^2_R}{2
k_B T_f}\right). \label{XII-Eq}
\end{eqnarray}
The ansatz (\ref{ansatzh}) provides the explicit link between the
space-velocity correlation moment and the parameters $\eta_i$:
$\langle {\bf r}\cdot{\bf v}_i\rangle_i=3\eta_i
(k_BT_f)^2/(m_i\omega_i^2)$. This expression in combination with
Eq. (\ref{XII-Eq}) permits one to derive the explicit expression
(\ref{tau'12-Eq}) for the relaxation time $\tilde{\tau}$ of the
space-velocity correlation equation.


\begin{thebibliography}{99}

\bibitem{fbmixture}
A. G. Truscott, K. E. Strecker, W. I. McAlexander, G. B.
Partridge, and R. G. Hulet, Science \textbf{291}, 2570 (2001); F.
Schreck, L. Khaykovich, K. L. Corwin, G. Ferrari, T. Bourdel, J.
Cubizolles, and C. Salomon, Phys. Rev. Lett. \textbf{87}, 080403,
(2001); Z. Hadzibabic, C. A. Stan, K. Dieckmann, S. Gupta, M. W.
Zwierlein, A. G\"orlitz, and W. Ketterle, {\it ibid} \textbf{88},
160401 (2002); G. Roati, F. Riboli, G. Modugno, and M. Inguscio,
{\it ibid} \textbf{89}, 150403 (2002); W. Hofstetter, J. I. Cirac,
P. Zoller, E. Demler, and M. D. Lukin, {\it ibid} \textbf{89},
220407 (2002); M. Greiner, C. A. Regal and D. S. Jin, Nature
(London) \textbf{426}, 537 (2003); M. W. Zwierlein, C. A. Stan, C.
H. Schunck, S. M. F. Raupach, A. J. Kerman, and W. Ketterle, Phys.
Rev. Lett. \textbf{92}, 120403 (2004); M. Greiner, C. A. Regal, C.
Ticknor, J. L. Bohn, and D. S. Jin, {\it ibid} \textbf{92}, 150405
(2004).


\bibitem{lahaye}
T.~Lahaye, Z.~Wang, G.~Reinaudi, S.~P.~Rath, J.~Dalibard, and
D.~Gu\'ery-Odelin, Phys. Rev. A {\bf 72}, 033411 (2005).

\bibitem{doyle}
Jonathan D. Weinstein, Robert deCarvalho, Cindy I. Hancox, and
John M. Doyle, Phys. Rev. A {\bf 65}, 021604(R), (2002).

\bibitem{monroe}
C. R. Monroe, E. A. Cornell, C. A. Sackett, C. J. Myatt, and C. E.
Wieman, Phys. Rev. Lett. \textbf{70}, 414 (1993).

\bibitem{arndt}
M. Arndt, M. Ben Dahan, D. Gu\'ery-Odelin, M. W. Reynolds, and J.
Dalibard, Phys. Rev. Lett. \textbf{79}, 625 (1997).

\bibitem{delannoy}
G. Delannoy, S. G. Murdoch, V. Boyer, V. Josse, P. Bouyer, and A.
Aspect, Phys. Rev. A \textbf{63}, 051602 (2001).

\bibitem{mosk}
A. Mosk, S. Kraft, M. Mudrich, K. Singer, W. Wohlleben, R. Grimm,
and M. Weidem\"uller, Appl. Phys. B \textbf{73}, 791 (2001).

\bibitem{rbk}
G. Ferrari, M. Inguscio, W. Jastrzebski, G. Modugno, G. Roati, and
A. Simoni, Phys. Rev. Lett. \textbf{89}, 053202 (2002).

\bibitem{schmidt}
P. O. Schmidt, S. Hensler, J. Werner, A. Griesmaier, A. G\"orlitz,
T. Pfau, and A. Simoni, Phys. Rev. Lett. \textbf{91}, 193201
(2003).

\bibitem{anderlini} M. Anderlini, E. Courtade, M. Cristiani, D. Cossart, D.
Ciampini, C. Sias, O. Morsch, and E. Arimondo, Phys. Rev. A
\textbf{71}, 061401 (2005).

\bibitem{kavoulakis}
G. M. Kavoulakis, C. J. Pethick, and H. Smith, Phys. Rev. A
\textbf{61}, 053603 (2000).

\bibitem{Nordheim}
L. W. Nordheim, Proc. Roy. Soc. London, \textbf{A 119}, 689
(1928).

\bibitem{Baym}
L.P. Kadanoff and G. Baym, {\it Quantum Statistical Mechanics}
 (W.A. Benjamin, New York, 1962).

\bibitem{vichi}
L. Vichi, J. Low Temp. {\bf 121}, 177 (2000).

\bibitem{joachain} J. C. Joachain, \emph{Quantum Collision Theory},
North Holland, Amsterdam (1983).

\bibitem{jookthomas}
R. Legere and K. Gibble, Phys. Rev. Lett. textbf{81}, 5780 (1998);
N. R. Thomas, N. Kj\ae rgaard, Paul S. Julienne, and Andrew C.
Wilson, {\it ibid.} \textbf{93}, 173201 (2004); Ch. Buggle, J.
L\'eonard,W. von Klitzing, and J.T.M.Walraven, {\it ibid.}
\textbf{93}, 173202 (2004).




\bibitem{dgoPRA99} D. Gu\'{e}ry-Odelin, F. Zambelli, J. Dalibard, and S.
Stringari, Phys. Rev. A \textbf{60}, 4851 (1999).

\bibitem{raizen}
T. P. Meyrath, F. Schreck, J. L. Hanssen, C.-S. Chuu, and M. G.
Raizen Phys. Rev. A \textbf{71}, 041604 (2005).


\bibitem{servaas} The phase shifts for $s$, $p$, $d$, $f$ and $g$ partial
 waves as a function of the collision wave vector $k$ have been
 provided by S. Kokkelmans (Private Communication).

\bibitem{walraven}
Ch. Buggle, P. Pedri, W. von Klitzing, and J. T. M. Walraven,
Phys. Rev. A {\bf 72}, 043610 (2005).

\bibitem{cercignani88} C. Cercignani, \emph{The Boltzmann equation and
its applications}, (Springer Verlag, New York, 1988).

\bibitem{boltzmann09} L. Boltzmann, in \emph{Wissenschaftliche
Abhandlungen}, edited by F. Hasenorl (J.A. Barth, Leipzig, 1909),
Vol II, p. 83.


\bibitem{numsim} G. A. Bird, \emph{Molecular Gas Dynamics and the Direct Simulation
of Gas Flows}, Clarendon, Oxford, (1994); H. Wu and C. J. Foot, J.
Phys. B \textbf{29}, L321 (1996); E. Cerboneschi, C. Menchini, and
E. Arimondo, Phys. Rev. A \textbf{62}, 013606 (2000).


\end{thebibliography}
\end{document}